\documentclass[twocolumn,showpacs,preprintnumbers,amsmath,amssymb,superscriptaddress,floatfix]{revtex4-1}

\usepackage{graphicx}
\usepackage{dcolumn}
\usepackage{bm}
\usepackage{tabularx}
\usepackage{multirow}
\usepackage{makecell}
\usepackage{braket}
\usepackage{ulem}
\usepackage{amsmath,amssymb} 
\newcolumntype{M}{>{\centering\arraybackslash}m{1.85cm}}
\usepackage[export]{adjustbox}
\usepackage{longtable}
\usepackage{float}
\usepackage{xcolor}
\usepackage{color}   
\usepackage{multirow}
\usepackage{hyperref}
\hypersetup{colorlinks=true, citecolor=blue, urlcolor=blue, linkcolor=blue}

\definecolor{mycolor}{rgb}{0.3, 0.2, 0.9}
\usepackage{lineno}

\begin{document}

\title{The measurement of the $^{99}$Tc $\beta$-decay spectrum and its implications for the effective value of weak axial coupling}

\author{J.~W. Song}
\affiliation{Korea Research Institute of Standards and Science, Daejeon 34113, South Korea}
\affiliation{Department of Physics, Kyungpook National University, Daegu 41566, South Korea}

\author{M.~Ramalho}
\affiliation{School of Physics, Engineering and Technology, University of York, Heslington, York YO10 5DD, United Kingdom}
\affiliation{Department of Physics, University of Jyv\"askyl\"a, P. O. Box 35 (YFL), FI-40014, Jyv\"askyl\"a, Finland}

\author{M.~K. Lee}
\affiliation{Korea Research Institute of Standards and Science, Daejeon 34113, South Korea}

\author{G.~B. Kim}
\affiliation{Lawrence Livermore National Laboratory, Livermore, CA 94550, USA}

\author{I. Kim}
\affiliation{Lawrence Livermore National Laboratory, Livermore, CA 94550, USA}

\author{H.~L. Kim}
\affiliation{Institute for Basic Science, Daejeon 34126, South Korea}

\author{Y.~C. Lee}
\affiliation{Institute for Basic Science, Daejeon 34126, South Korea}
\affiliation{Department of Physics and Astronomy, Seoul National University, Seoul 08826, South Korea}

\author{K.~R. Woo}
\affiliation{Institute for Basic Science, Daejeon 34126, South Korea}

\author{J.~Kotila}
\affiliation{Finnish Institute for Educational Research, University of Jyv\"askyl\"a, Jyv\"askyl\"a, Finland}

\author{J.~Kostensalo}
\affiliation{Natural Resources Institute Finland, Yliopistokatu 6B, FI-80100 Joensuu, Finland}

\author{J.~Suhonen}~\email{jouni.t.suhonen@jyu.fi}
\affiliation{Department of Physics, University of Jyv\"askyl\"a, P. O. Box 35 (YFL), FI-40014, Jyv\"askyl\"a, Finland}
\affiliation{International Centre for Advanced Training and Research in Physics (CIFRA), 077125 Bucharest-Magurele, Romania}

\author{H.~J.~Kim}~\email{hongjoo@knu.ac.kr}
\affiliation{Department of Physics, Kyungpook National University, Daegu 41566, South Korea}

\date{\today}

\begin{abstract}
Measurements of $\beta$-spectral shapes is an important way to examine the effective value of the weak axial coupling $g_{\rm A}$. These studies focus specifically on forbidden non-unique $\beta^-$ transitions, as only in these cases is the spectral shape directly sensitive to the ratio $g_{\rm A}/g_{\rm V}$. Here, the value of the weak vector coupling constant,  $g_{\rm V}$, is fixed at 1.0 according to the Conserved Vector Current (CVC) hypothesis. In previous studies for the fourth-forbidden non-unique $\beta^-$ decays of $^{113}$Cd 
[J.~Kostensalo \textit{et al.}, Phys. Lett. B 822, 136652 (2021)] and $^{115}$In [A.~F. Leder \textit{et al.}, Phys. Rev. Lett. 129, 232502 (2022) and L. Pagnanini \textit{et al.}, Phys. Rev. Lett. 133, 122501 (2024)] a quenched value was determined for the ratio 
$g_{\rm A}/g_{\rm V}$ using $g_{\rm V}=1.0$. A notable exception is the recent measurement and analysis of the second-forbidden non-unique $\beta$-decay transition in $^{99}$Tc, performed by M. Paulsen \textit{et al.},  Phys. Rev. C 110, 05503(2024). Where an enhanced ratio $g_{\rm A}/g_{\rm V}=1.526(92)$ was suggested. To resolve this apparently contradictory situation with the effective value of $g_{\rm A}$, we have performed calculations based on the nuclear shell model (NSM) Hamiltonians glekpn, jj45pnb, and the MQPM approach with a careful consideration of the small relativistic vector nuclear matrix element (sNME). The theoretical spectra were compared to the $^{99}$Tc $\beta$-decay spectrum by using the 4$\pi$ gold absorber with a Metallic Magnetic Calorimeter (MMC). In all cases, we found that the data matches well with reduced $g_{\rm A}$/$g_{\rm V}$ values of 1.0--1.2. Our result contradicts the previously reported measurement for $^{99}$Tc and instead supports a quenched axial coupling as reported for other isotopes.
\end{abstract}

\maketitle

\textit{Introduction.}---Electrons and electron antineutrinos are emitted in $\beta^-$ decays as a result of a transmutation of a parent nucleus $(Z,N)$ to a daughter nucleus $(Z+1,N-1)$, $Z$ being the number of protons and $N$ the number of neutrons. Studies of $\beta$-electrons (electrons emitted in $\beta^-$ decays) have become an important subject of study, e.g., in connection with nuclear reactors and the anomalies in their fluxes of antineutrinos, and rare-events experiments searching for beyond-the-standard-model physics, like those trying to measure rare $\beta$-decays and double beta decays. The $\beta$-electrons can also be severe backgrounds in these experiments. In all the aforementioned cases, measurements of the associated $\beta$-spectral shape is essential. Distortions in $\beta$-decay spectral shapes also provide important input for understanding the weak axial coupling constant $g_{\rm A}$, which is relevant for sensitivity estimates in current and future beyond-the-standard-model experiments; see the reviews~\cite{Eng2017, Eji2019, Ago2023}.

In allowed $\beta$ decays, such as Fermi and Gamow--Teller transitions, the emitted lepton pair (electron and antineutrino) carries zero orbital angular momentum~\cite{Suh2007}. In contrast, forbidden decays occur when the emitted lepton pair carries nonzero orbital angular momentum~\cite{Beh1982}. The degree of forbiddenness can range from first-forbidden to $n$:th-forbidden, with each additional level typically increasing the decay half-life by a factor of approximately $10^4$. 
Forbidden $\beta$-transitions are generally classified into two types: unique and non-unique. Unique transitions are characterized by a single dominant nuclear matrix element~(NME) and thus exhibit a universal $\beta$-spectral shape that is independent of nuclear structure~\cite{Suh2007}. In contrast, non-unique transitions involve a complex interplay of several NMEs, making the $\beta$-decay spectrum highly sensitive to the nuclear wave functions of both the initial and final states~\cite{Beh1982, Haa2016, Haa2017}.

The (partial) half-life of a forbidden non-unique $\beta$-transitions also depends on the values of the weak vector and axial-vector couplings, $g_{\rm V}$ and $g_{\rm A}$, respectively~\cite{Beh1982, Haa2016, Haa2017}. The conserved vector current (CVC) hypothesis sets $g_{\rm V}=1.0$ whereas the partially conserved axial-vector current (PCAC) hypothesis leads to an effective (quenched) value of $g_{\rm A}^{\rm eff}$ in finite nuclei~\cite{Suh2017}, its bare-nucleon value $g_{\rm A}=1.27$ resulting from the experimental studies of the decay of an isolated neutron. The quenching of the effective value of $g_{\rm A}$ has recently been studied extensively in~\cite{Eji2019, Suh2017, Suh2019}, and in particular its effects on the neutrinoless double beta decay have been addressed in~\cite{Eng2017, Ago2023}. To quantify the quenched value of 
$g_{\rm A}$, combined experimental and theoretical analyses of $\beta$-spectral shapes of individual $\beta^-$ transitions have recently been performed for the fourth-forbidden non-unique $\beta$-decays of $^{113}$Cd~\cite{Bod2020, Kos2021, Kos2023, Ban2024} and $^{115}$In~\cite{Led2022, Pag2024, Kos2024}. 

A recent measurement of the $\beta$-decay spectrum for the second-forbidden non-unique ground-state-to-ground-state transition $^{99}\textrm{Tc}(9/2^+) \to\,^{99}\textrm{Ru}(5/2^+)$, which has a 100\% branching ratio, was reported by Paulsen \textit{et al.}~\cite{Pau2023}. The accompanying theoretical analysis yielded a contradictory result, suggesting an enhancement of the axial coupling with a value of $g_{\rm A}/g_{\rm V} = 1.526(92)$ using $g_{\rm V} = 1.0$.

The gA-EXPERiment and Theory (gA-EXPERT) collaboration is an international effort dedicated to quantifying the effective axial-vector coupling constant, $g_{\rm A}^{\rm eff}$, by precisely measuring $\beta$-decays of various nuclei and comparing the results with state-of-the-art nuclear theory calculations.
To address the tension raised by the enhanced value of $g_{\rm A}^{\rm eff} = 1.526(92)$ reported in Ref.~\cite{Pau2023}, we performed a high-precision measurement of the $^{99}$Tc $\beta$-decay spectrum using metallic magnetic calorimeter quantum sensors with a low-energy threshold of 10~keV. The measured spectrum was compared to predictions from three nuclear models, each based on distinct nuclear Hamiltonians. In this Letter, we present a quantitative comparison between data and modern nuclear theory, providing new constraints on $g_{\rm A}^{\rm eff}$ and demonstrating the sensitivity of precision $\beta$-spectroscopy to nuclear structure effects.

\textit{Theory.}---To first order, the half-life of a $\beta$-decay transition can be written as:
\begin{equation}
t_{1/2}=\frac{\kappa}{\tilde{C}} \,,
\end{equation}
where $\kappa$ is a constant~\cite{Kum2020,Kum2021} and $\tilde{C}$ is the integrated shape function $C$, which incorporates various phase-space factors and NMEs arising in the next-to-leading-order expansion, as discussed in detail in Refs.~\cite{Haa2016,Haa2017}. Our $\beta$-spectral shape and half-life analyses employ screening, radiative, and atomic-exchange corrections. For electron energies below some tens of keV, the distortion in the $\beta$-decay spectral shape is mostly dominated by the atomic exchange correction, as originally studied for allowed $\beta$-decays~\cite{Nitescu2023}. This correction is responsible for the sharp upward trend of the spectral shape at low energies.

The complexity of the shape function $C$ can, however, be cast in a very simple dependence on the weak couplings by writing
\begin{equation}
C(w_e) = g_{\rm V}^2C_{\rm V}(w_e) + g_{\rm A}^2C_{\rm A}(w_e) + g_{\rm V}g_{\rm A}C_{\rm VA}(w_e) \,
\end{equation}
where $w_e$ is the total (rest-mass plus kinetic) energy of the emitted electron. 
The shape of the $\beta$-electron spectrum is sensitive to the value of $g_{\rm A}$ due to subtle interference effects among the vector $C_{\rm V}(w_e)$, axial-vector $C_{\rm A}(w)$, and mixed vector–axial-vector $C_{\rm VA}(w)$ contributions to the transition matrix element~\cite{Haa2016}. For some nuclei (very few, as charted up to present days~\cite{Eji2019, Kum2020, Kum2021, Ram2024, Ram2024c}) this causes measurable changes in the $\beta$-spectra. Thus, the effective value $g_{\rm A}^{\rm eff}$ can be found by computing theoretical $\beta$-spectra for a set of $g_{\rm A}^{\rm eff}$ values and comparing them with the measured one. This method, coined as the spectrum-shape method~(SSM), was originally proposed in~\cite{Haa2016}.

Another factor influencing theoretical analyses of $\beta$-spectral shapes and half-lives is the so-called small relativistic vector nuclear matrix element~(sNME)~\cite{Beh1982}. The sNME gathers contributions outside the valence major shell that contains the proton and neutron Fermi surfaces, which makes its calculation particularly hard for nuclear-theory frameworks used, e.g., in the present work and also in the work of Paulsen \textit{et al.}~\cite{Pau2023}. Despite its smallness, the sNME can influence the $\beta$-spectral shapes and half-lives quite strongly~\cite{Kos2021, Kos2023, Kum2021, Ram2024}. 

A realistic value of the sNME can be estimated from its  CVC value~\cite{Beh1982}, first used for spectral-shape calculations in Ref.~\cite{Kum2020} and later by Paulsen \textit{et al.}~\cite{Pau2023}. It should be noted, though, that the CVC value represents an  ``ideal'', pertaining to a perfect many-body theory, which is not the case for the nuclear-structure frameworks typically used for $\beta$-decay calculations of heavy open-shell nuclei, like the presently discussed $^{99}$Tc. To circumvent the use of the CVC value, the experimental works~\cite{Kos2021,Ban2024,Pag2024} and theoretical works~\cite{Kos2023, Kos2024} used sNME as a fitting parameter, together with $g_{\rm A}^{\rm eff}$. By this, it was possible to reproduce both the measured $\beta$-spectral shape and the experimental (partial) half-life of the corresponding $\beta$-transition without losing the CVC value $g_{\rm A}=1.0$ for the vector coupling. This approach, termed the enhanced spectrum-shape method (ESSM) in Ref.~\cite{Kum2021}, is adopted in the present work.
 
In this Letter, we present new theoretical calculations of the $\beta$ spectrum of $^{99}$Tc (with $Z = 43$ and $N = 56$) and compare them to the measured spectrum in order to extract the $g_{\rm A}^{\rm eff}$, using ESSM.
To gain preliminary insights into the effective axial coupling $g_{\rm A}^{\rm eff}$ in the $\beta$-decay of $^{99}$Tc, we performed nuclear shell model (NSM) calculations and compared the results with the measured spectrum reported in Ref.~\cite{Pau2023}. As the $\beta$-decay rate has a quadratic dependence on the sNME, two solutions were found for each value of $g_{\rm A}^{\rm eff}$, one closer to [sNME(c)] and one further away from [sNME(f)] CVC value. In references~\cite{Ram2024,Ram2024b} the related two sets of theoretical $\beta$-spectra of $^{99}$Tc were introduced for two NSM Hamiltonians, namely \textit{jj45pnb}~\cite{Lis2004} and \textit{glekpn}~\cite{Mac1990}; In that preliminary study, for the \textit{glekpn} interaction we've used a set of single particle energies fit to the $A=94-98$ mass region, however here we use the default single particle energies since it yields better experimental agreement for this isotope.
Comparison with the measured spectrum~\cite{Pau2023} indicated that both sNME values used with the \textit{jj45pnb} Hamiltonian yielded good agreement, while the fit is somewhat poorer with sNME(c), and significantly worse with sNME(f) when using the \textit{glekpn} Hamiltonian, further details of this preliminary study and its calculations can be found in~Refs~\cite{Ram2024b, Ram2024}.

In addition to the two NSM Hamiltonians described above, we also employ the nuclear many-body method known as the Microscopic Quasiparticle-Phonon Model (MQPM). A concise presentation of the essentials of NSM, and MQPM can be found in Ref.~\cite{Eji2019}. 
It should be pointed out here that Ref.~\cite{Pau2023} employed the NSM in their analysis, adopting the CVC value for the sNME without fitting it to the experimental half-life of $\beta$-decay. A systematic comparison of various theoretical models with the measured spectrum offers a more robust framework for extracting the $g_{\rm A}^{\rm eff}$ from high-precision spectral data.

\textit{Experiment.}--- Metallic magnetic calorimeters (MMCs) are some of the most advanced cryogenic detectors for precision energy measurements. These devices utilize paramagnetic ions embedded in metallic hosts—typically Au:Er or Ag:Er—as temperature sensors thermally coupled to metallic absorbers. The paramagnetic material is placed in a weak magnetic field ($<$10~mT) and operated at temperatures below 100~mK. 
When an incident particle deposits energy $E$ in the absorber, it induces a temperature rise $\Delta T$ in the paramagnetic bath. This temperature increase leads to a decrease in the total magnetization of the paramagnetic ions, which is sensed by a superconducting pickup coil. The magnetic signal is then transmitted to a superconducting quantum interference device (SQUID) for precise readout.
In this work, we employed an MMC fabricated at the Korea Research Institute of Standards and Science (KRISS), incorporating 420~ppm of Ag:$^{168}$Er as the sensor material~Refs~\cite{MMC2024, AgEr2025}.

Radionuclide sources were prepared in the form of $4\pi$ solid-angle absorbers. A solution of $^{99}$Tc dissolved in nitric acid was dropped onto a gold foil (2~mm~$\times$~2~mm $\times$~33~\textmu m) and subsequently dried.
The foil, containing approximately 10~Bq of $^{99}$Tc activity, was then folded and rolled three times in a process known as the kneading method~\cite{kneading2015, kneading2016}. This technique enhances the uniformity of $^{99}$Tc distribution within the absorber and improves energy resolution.
A portion of the kneaded foil was cut, encapsulated between two 25~\textmu m thick gold foils, and compressed using a rolling machine. The final encapsulated sample, measuring 1.2~mm~$\times$~1.3~mm~$\times$~50~\textmu m, was thermally coupled to the MMC sensor. 
The total absorber mass was 1.75~mg, and the total heat capacity, including the contribution from the paramagnetic ions in the MMC sensor, was approximately 400~pJ in the 15-50~mK temperature range. The final $^{99}$Tc activity was below 1~Bq.

\begin{figure}[hbt!] 
    \centering 
    \includegraphics[width=0.5\textwidth]{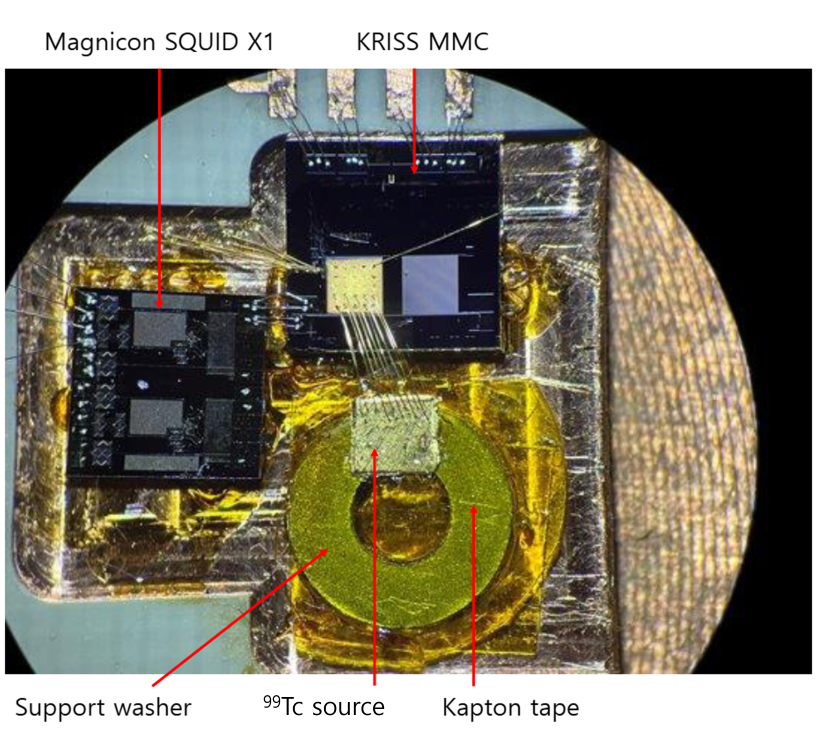} 
    \caption{The detector setup used for the $\beta$-decay measurement. The gold sample containing the $^{99}$Tc source is thermally coupled to the MMC sensor via gold wire bonds. A support washer with a Kapton tape on top for thermal insulation was placed below the sample to align its height with that of the MMC sensor.} 
    \label{fig:MMC-SQUID} 
\end{figure}

The detector setup consists of an MMC and a Magnicon SQUID~(Model~X1) for signal readout (see Figure~\ref{fig:MMC-SQUID}). The detector was installed and operated in a pulsed tube pre-cooled $^{3}$He/$^{4}$He dilution refrigerator at Lawrence Livermore National Laboratory (LLNL). The signal was amplified using an additional array SQUID (Model X16) installed at the 4~K stage.

The output was low-pass filtered using a Stanford Research Systems SRS 560 amplifier~\cite{SRS560} with a 30~kHz cutoff frequency. Data acquisition was performed over 34 hours at 15~mK in un-triggered continuous mode to capture the full information, with a sampling rate of 200~kS/s. A software trigger was applied using a trapezoidal filter with a 1-ms shaping time.

For energy calibration, an external $^{133}$Ba source was used as a reference, while a $^{148}$Gd source was introduced inside the setup for internal calibration. 
Although the $\alpha$ signals from $^{148}$Gd were not observed due to slight misalignment, characteristic X-rays from $^{148}$Gd interactions were successfully detected. The simultaneous observation of $\gamma$ rays from $^{133}$Ba and X-rays resulting from $^{148}$Gd provided multiple reference points for accurate calibration of the energy scale.

Two trapezoidal filters with different shaping parameters were applied to the raw pulses. The long trapezoid was used to obtain information about the entire DC signal, while the short trapezoid was designed to capture information corresponding only to the rise time of the DC signal~\cite{trap1994, pile2020}. All data were analyzed using custom MATLAB$^{\scalebox{0.7}{\textcircled{R}}}$ scripts~\cite{mat2016b}. 

\textit{Results.}---In the background dataset, a total of 130,001 events were recorded. Among them, 80,903 events exceeded the energy threshold defined for the analysis as 10 keV. A pulse shape discrimination (PSD) cut was subsequently applied to select events consistent with $^{99}$Tc signal-like pulses. After that, 77,487 events remained for further analysis.

Similarly, the calibration dataset obtained with an external $^{133}$Ba source contained 47,491 events. Out of these, 35,821 events passed the energy threshold. After applying the same PSD criteria, 23,596 events were retained.

\begin{figure}[hbt!] 
    \centering 
    \includegraphics[width=\columnwidth]{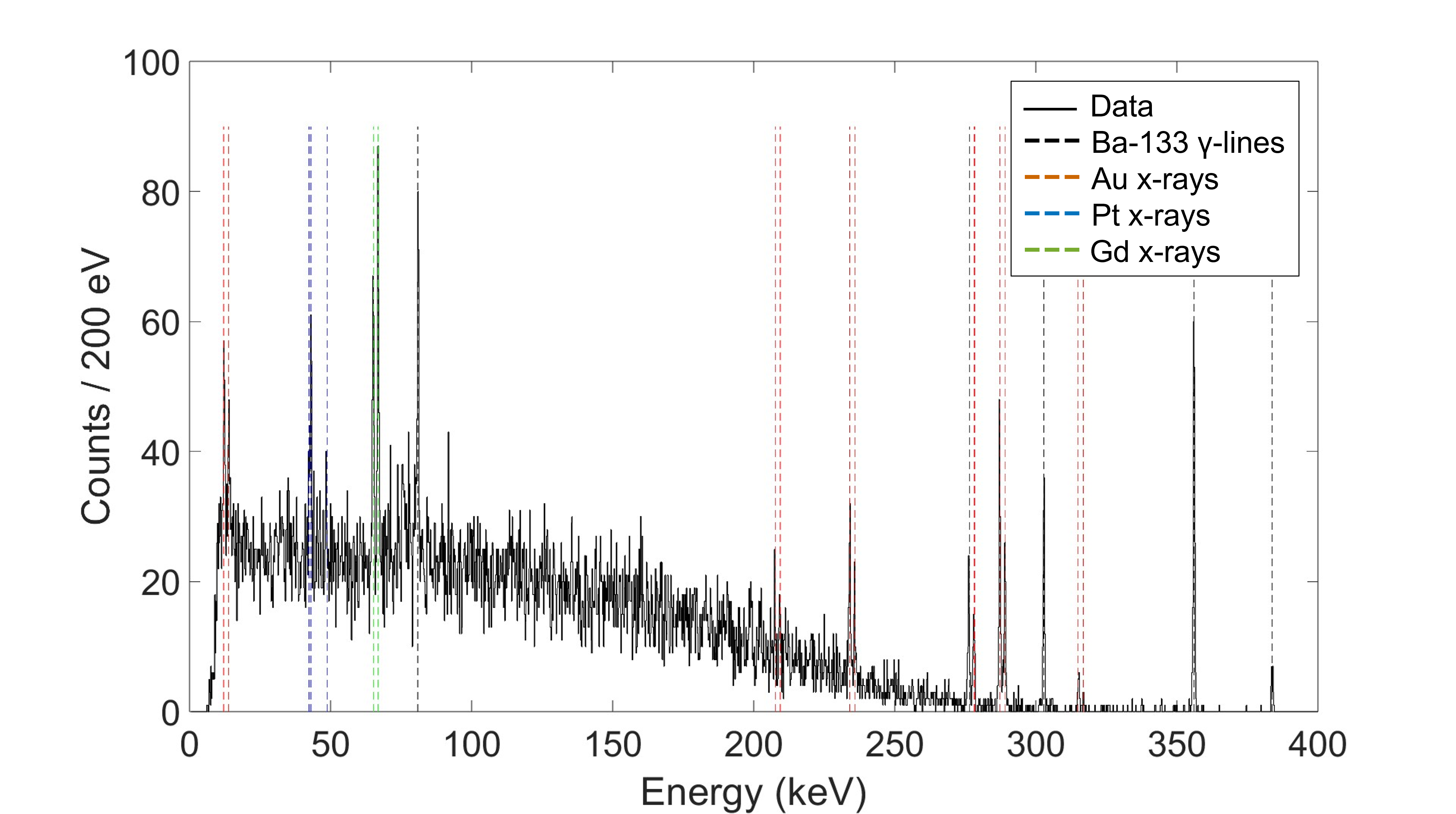} 
    \caption{$^{99}$Tc spectrum with external $^{133}$Ba calibration source. Details of the X-ray peaks are provided in Table~\ref{Table:Xray_Peaks}.} 
    \label{fig:Calibration} 
\end{figure}

Figure~\ref{fig:Calibration} presents the $^{99}$Tc spectrum measured with the external $^{133}$Ba source in place. Characteristic X-ray peaks from Gd and Pt fluorescence, as well as $\gamma$ lines from $^{133}$Ba, are clearly visible~\cite{xray2003}. These peaks, along with the $\gamma$ lines from $^{133}$Ba and escape peaks due to partial energy deposition, were used for energy calibration~\cite{khazov2011nds}. The peak positions were determined by Gaussian fitting with a linear background model, and the results are summarized in Table~\ref{Table:Xray_Peaks}.

Most peaks exhibit excellent agreement with literature values, with $\Delta E$ typically below 0.3~keV and energy resolutions better than 1~keV in the 66–383~keV energy range. The entries in Table~\ref{Table:Xray_Peaks} labeled as “Not distinguished,” including K$_{\beta_3}$ and K$_{\beta_1}$ transitions from Gd X-ray and Au escape lines, correspond to peaks with low relative intensities~\cite{LLBL2009x}. Given the limited event statistics, these peaks could not be individually resolved in the spectral fit. The uncertainties reported for each value reflect only the statistical errors derived from the Gaussian fits, without accounting for potential systematic contributions.

The Gd X-ray peaks from the internal $^{148}$Gd source and the $^{133}$Ba $\gamma$ lines were used as primary reference points for the calibration. Additional fluorescence lines from Pt are attributed to materials in the detector setup and collimator. Several low-energy escape peaks (below 20~keV) show broader FWHM and degraded relative resolution due to increased noise and limited event statistics.

\begin{table*}[hbt!]
\centering
\caption{Measured X-ray and gamma-ray lines, including their tabulated and measured energies.}
\resizebox{\textwidth}{!}{
\begin{tabular}{ c c c c c c c}
\hline\hline
\textbf{Origin} & \multicolumn{2}{c}{\textbf{Radiation type}} 
&\textbf{\begin{tabular}[c]{@{}c@{}} Tabulated \\ energy {[}keV{]}~\cite{xray2003, khazov2011nds}\end{tabular}}
&\textbf{\begin{tabular}[c]{@{}c@{}} Measured \\ energy {[}keV{]}\end{tabular}}
&\textbf{\begin{tabular}[c]{@{}c@{}} FWHM \\ resolution {[}keV{]}\end{tabular}}
& \textbf{$\Delta$E [keV]} \\
\hline

\multirow{4}{*}{\begin{tabular}[c]{@{}c@{}}Gd fluorescence \\ from internal source\end{tabular}}
 & \multicolumn{2}{c}{X-ray K$_{\alpha_{2}}$} & 42.3097(13) & 42.30(10) & 0.9(4) & 0.01(10)\\
 & \multicolumn{2}{c}{X-ray K$_{\alpha_{1}}$} & 42.9968(12) & 42.9427(31) & 0.686(10) & 0.0541(33)\\ 
 & \multicolumn{2}{c}{X-ray K$_{\beta_{3}}$} & 48.5505(21) & \multirow{2}{*}{48.29(5)\footnotemark[1]} & \multirow{2}{*}{1.05(22)} & \multirow{2}{*}{0.41(5)} \\ 
 & \multicolumn{2}{c}{X-ray K$_{\beta_{1}}$} & 48.6957(21) \\ 
 \hline

\multirow{2}{*}{\begin{tabular}[c]{@{}c@{}}Pt fluorescence \\ from collimator\end{tabular}} 
 & \multicolumn{2}{c}{X-ray K$_{\alpha_{2}}$} & 65.1226(21) & 65.005(22) & 0.96(6) & 0.118(22) \\
 & \multicolumn{2}{c}{X-ray K$_{\alpha_{1}}$} & 66.8311(20) & 66.78(5) & 0.65(15) & 0.06(5) \\
 \hline

\multirow{5}{*}{$^{133}$Ba calibration lines} 
 & \multicolumn{2}{c}{$\gamma$} & 80.9979(11)  & 80.881(24) & 0.77(8) & 0.117(25) \\
 & \multicolumn{2}{c}{$\gamma$} & 276.3989(12) & 276.12(5) & 0.57(13) & 0.28(5) \\
 & \multicolumn{2}{c}{$\gamma$} & 302.8508(5) & 302.764(21) & 0.60(5)  & 0.087(21) \\
 & \multicolumn{2}{c}{$\gamma$} & 356.0129(7) & 355.926(9) & 0.653(21) & 0.087(9) \\
 & \multicolumn{2}{c}{$\gamma$} & 383.8485(12) & 383.69(6) & 0.82(14) & 0.16(6) \\
 \hline

\multirow{18}{*}{Escape lines}

\\

& \textbf{\begin{tabular}[c]{@{}c@{}} Absorbed  \\ $^{133}$Ba $\gamma$ line [keV] \end{tabular}}

& \textbf{\begin{tabular}[c]{@{}c@{}} Escaping  \\ Au X-ray [keV] \end{tabular}} & & & &\\
\hline

 & \multirow{2}{*}{383.8485} 
 & 66.993(23) {[}K$_{\alpha_{2}}${]} & 316.8555(26) & 316.817(18) & 0.37(5) & 0.038(18) \\
 && 68.8069(22) {[}K$_{\alpha_{1}}${]} & 315.0416(25) & 315.09(4) & 0.72(9) & 0.05(4) \\
 \cline{2-7}

 & \multirow{4}{*}{356.0129} 
 & 66.993(23) {[}K$_{\alpha_{2}}${]} & 289.0199(24) & 288.943(21) & 0.51(5) & 0.077(21) \\
 && 68.8069(22) {[}K$_{\alpha_{1}}${]} & 287.206(23) & 287.051(26) & 0.59(6) & 0.155(26) \\
 && 77.5773(35) {[}K$_{\beta_{3}}${]} & 278.4356(36) & \multirow{2}{*}{277.911(31)\footnotemark[1]} & \multirow{2}{*}{0.88(8)} & \multirow{2}{*}{0.122(31)}\\
 && 77.983(37) {[}K$_{\beta_{1}}${]}  & 278.0329(38)\\
 \cline{2-7}
 
 & \multirow{2}{*}{302.8508} 
 & 66.993(23) {[}K$_{\alpha_{2}}${]} & 235.8578(24) & 235.659(23) & 0.47(6) & 0.199(23) \\
 && 68.8069(22) {[}K$_{\alpha_{1}}${]} & 234.0439(23) & 233.99(4) & 0.63(11) & 0.05(4) \\
 \cline{2-7}
 
 & \multirow{2}{*}{276.3989} 
 & 66.993(23) {[}K$_{\alpha_{2}}${]} & 209.4059(26) & 209.137(12) & 0.72(4) & 0.269(12) \\
 && 68.8069(22) {[}K$_{\alpha_{1}}${]} & 207.592(25) & 207.36(5) & 0.62(13) & 0.24(5) \\
 \cline{2-7}
 
 & \multirow{2}{*}{80.9979}
 & 66.993(23) {[}K$_{\alpha_{2}}${]} & 14.0049(25) & 13.87(5) & 1.49(21) & 0.13(5) \\
 && 68.8069(22) {[}K$_{\alpha_{1}}${]} & 12.191(25) & 12.144(17) & 0.97(6) & 0.046(17) \\
\hline\hline
\footnotetext[1]{Not distinguished}
\end{tabular}}
\label{Table:Xray_Peaks}
\end{table*}

A linear energy calibration was performed using the five most prominent $\gamma$ peaks from $^{133}$Ba: 383.8~keV, 356.0~keV, 302.8~keV, 276.4~keV, and 80.9~keV. The resulting calibration function was then applied to the analysis dataset. The positions of the Gd X-ray peaks were compared between the calibration and analysis datasets, and no significant discrepancies were observed. This consistency validated the use of the same calibration function for both datasets. Specifically, the 42.3097~keV, 42.9968~keV, and 48.6957~keV peaks in the two datasets differed by 0.14(11)~keV, 0.093(18)~keV, and 0.38(5)~keV, respectively—well within the corresponding uncertainties.

\begin{figure}[hbt!] 
    \centering 
    \includegraphics[width=0.5\textwidth]{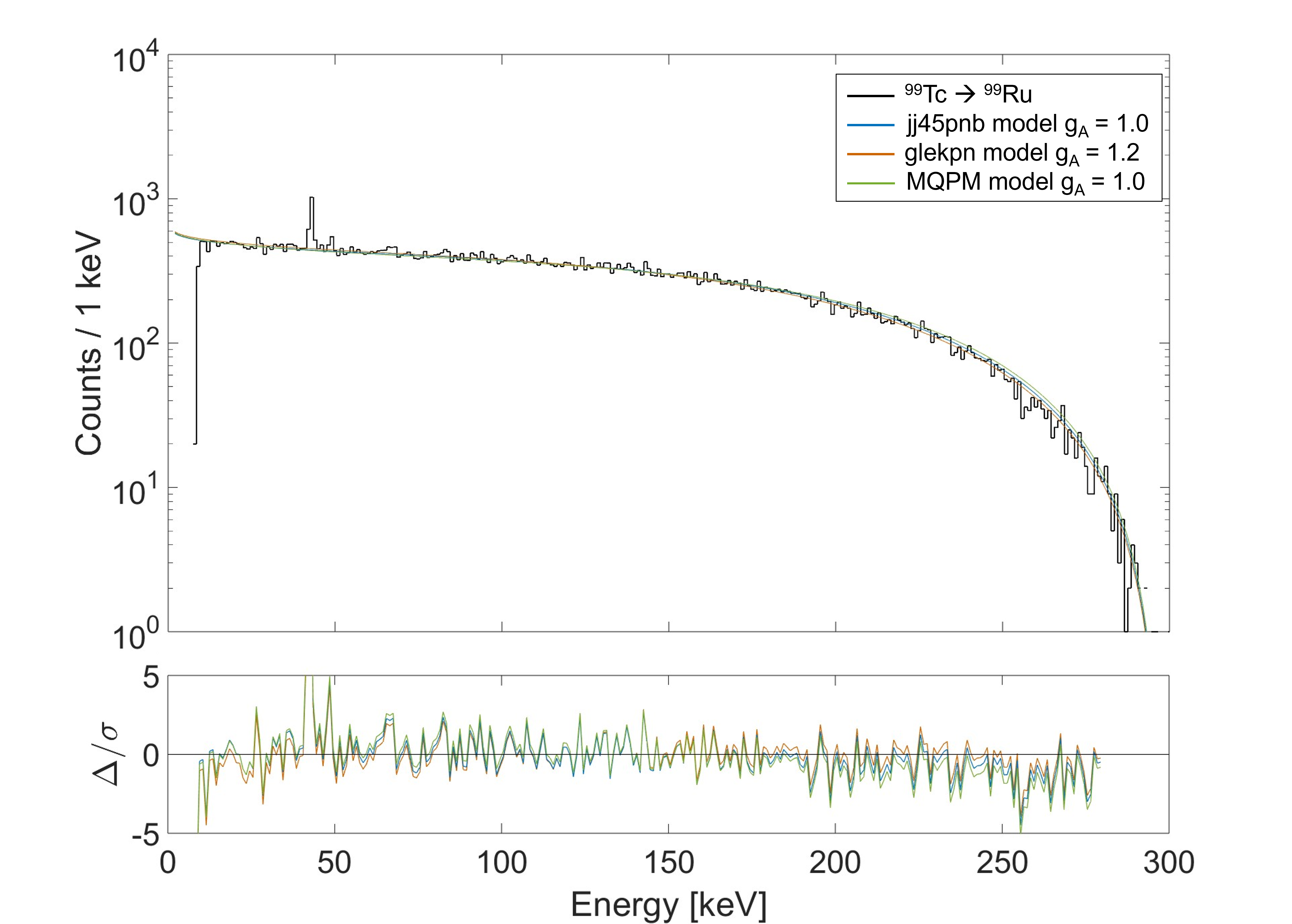} 
    \caption{Spectrum of calibrated $^{99}$Tc, plotted with a 1~keV binning, compared with theoretical predictions based on the farCVC variants of the \textit{glekpn}, \textit{jj45pnb}, and MQPM models. The models were fitted using a $\chi^2_{\nu}$ method. The residual graph below was calculated over the energy range from 0–280~keV, corresponding to the $\chi^2_{\nu}$ fitting range.
} 
    \label{fig:model_fit} 
\end{figure}
In our ESSM theory framework, templates of the $\beta$-spectra were generated for a set of $g_{\rm A}$ values (with $g_{\rm V}=1.0$), adjusting simultaneously the sNME value(s) to correctly reproduce the decay half-life. Then, reduced chi-square ($\chi^2_{\nu}$) fitting was performed using the provided models, applying simple shape comparisons and calculating residuals to identify the energy regions with the most significant deviations. The bin size for each spectrum was set to 1~keV. 

A reduced chi-squared ($\chi^2_{\nu}$) fitting procedure was then applied to compare the theoretical models with the measured spectrum. The bin size for each spectrum was set to 1~keV. The fits were performed by minimizing $\chi^2_{\nu}$ over the energy ranges 10–41~keV and 50–280~keV, excluding the Gd X-ray region. Simple shape comparisons and residual analyses were used to identify energy regions with significant deviations. Figure~\ref{fig:model_fit} visually demonstrates how well each model aligns with the $^{99}$Tc data. 

\begin{table}[hbt!]
\centering
\caption{Results for the three considered nuclear models on the parameters of interest $g_{\rm A}$. The reduced chi-square and p-value are also reported as an estimator of the goodness of fit.}
\footnotesize
\begin{tabular}{l c c c c c }
\hline\hline
\textbf{Model}   & \textbf{$g_{\rm A}^{\rm eff}$} & sNME [fm$^3$] & \textbf{$\chi ^2_{\nu}$} & \textbf{$p$-value}  \\ \hline
CloseCVC         &                &            &                       &                  \\
\textbf{glekpn}  & 1.1            & 0.0674           & 1.1057                & 0.12           \\
\textbf{jj45pnb} & 1.0            & 0.0681           & 1.2436                & 0.004           \\
\textbf{MQPM}    & 1.0            & 0.0651           & 1.6437                & $<$0.0001           \\
\\
FarCVC (Best fit)           &                &            &                       &                  \\
\textbf{glekpn}  & 1.2            & -0.0698            & 1.0664                & 0.22           \\
\textbf{jj45pnb} & 1.0            & -0.0669           & 1.0966                & 0.14           \\
\textbf{MQPM}    & 1.0            & -0.0683            & 1.0951                & 0.14           \\ 
\hline\hline
\end{tabular}
\label{Table:results}
\end{table}
We performed the fitting using $g_{\rm A}^{\rm eff}$ with both the closeCVC [sNME(c)] and farCVC [sNME(f)] models, and the results are summarized in Table~\ref{Table:results}. For the farCVC cases, the \textit{glekpn} model at $g_{\rm A}^{\rm eff}$ = 1.2 yields the best agreement with the experimental spectrum, with a reduced chi-squared value of $\chi^2_{\nu}$ = 1.0664 and a corresponding $p$-value of 0.22. The \textit{jj45pnb} model at $g_{\rm A}^{\rm eff}$ = 1.0 also provides a good match with $\chi^2_{\nu}$ = 1.0966 and $p$-value of 0.14. Similarly, the MQPM model at $g_{\rm A}^{\rm eff}$ = 1.0 gives $\chi^2_{\nu}$ = 1.0951 and $p$-value of 0.14. All fits were performed with $\nu$ = 261 degrees of freedom.

The reduced chi-squared values close to unity and $p$-values well above 0.05 indicate that all three farCVC models are statistically compatible with the measured spectrum. This consistency supports the reliability of the farCVC scenario, where the sNME is adjusted to reproduce the experimental half-life. Among the models considered, \textit{glekpn} slightly outperforms the others, but all three suggest that a quenched axial-vector coupling of $g_{\rm A}^{\rm eff}$ = 1.0--1.2 provides the best agreement with the data.
Our result shows a clear deviation from that reported in Ref.~\cite{Pau2023}, and instead exhibits better consistency with a quenched $g_{\rm A}$ value comparable to those extracted from other nuclei.

\textit{Conclusions.}---In this work, we performed a high-resolution measurement of the $\beta$-decay spectrum of $^{99}$Tc using metallic magnetic calorimeter quantum sensors and compared the results with predictions from multiple nuclear models. The extracted value of the effective axial-vector coupling, $g_{\rm A}^{\rm eff}$, deviates from the enhanced value reported in Ref.~\cite{Pau2023} and is more consistent with values observed in other nuclei.
We attribute this discrepancy primarily to the different treatment of the sNME adopted in Ref.~\cite{Pau2023}. By applying the ESSM, which simultaneously fits the measured $\beta$-spectrum and reproduces the partial half-life, we placed more robust constraints on $g_{\rm A}^{\rm eff}$ based on three nuclear theory approaches.

Our results for the second-forbidden non-unique $\beta^-$ decay of $^{99}$Tc provide valuable insight into the effective axial-vector coupling strength $g_{\rm A}^{\rm eff}$ in the mass region. The extracted quenching pattern is consistent with those observed in heavier nuclei, such as $^{113}$Cd~\cite{Bod2020, Kos2021, Kos2023, Ban2024} and $^{115}$In~\cite{Led2022, Pag2024, Kos2024}, suggesting that the quenching of $g_{\rm A}$ may follow a systematic trend across mass regions. This implies that the same level of quenching could be applied to both allowed and forbidden virtual transitions contributing to $0\nu\beta\beta$ NMEs, thereby simplifying the theoretical evaluation of NMEs.

In particular, our findings can inform the modeling of $0\nu\beta\beta$ decay in $^{100}$Mo, which lies in the same mass region as $^{99}$Tc. If the observed quenching behavior persists across neighboring nuclei, it could facilitate a more unified and robust framework for $0\nu\beta\beta$ nuclear matrix element calculations in this mass region.

These results demonstrate the sensitivity of precision $\beta$-spectroscopy to nuclear structure effects and provide valuable benchmarks for the development and validation of nuclear theory—an effort central to the mission of the gA-EXPERT collaboration.

\textit{Acknowledgments.}---  H. J. Kim acknowledges support by the National Research Foundation of Korea (NRF) grant, funded by the Korean government (MSIT), Contract No. RS-2024-00348317. J. W. Song acknowledges support by the National Research Foundation of Korea (NRF) grant, funded by the Ministry of Education, Contract No. RS-2023-00274589. J. Kotila and J. Suhonen acknowledge support from project PNRR-I8/C9-CF264, Contract No. 760100/23.05.2023 of the Romanian Ministry of Research, Innovation and Digitization (the NEPTUN project). M. Ramalho acknowledges support by the Oskar Huttunen Foundation. We acknowledge grants of computer capacity from the Finnish Grid and Cloud Infrastructure (persistent identifier urn:nbn: fi:research-infras-2016072533) and the support by CSC–IT Center for Science, Finland, for generous computational resources. G. B. Kim and I. Kim acknowledge support from the U.S. Department of Energy, performed under the auspices of Lawrence Livermore National Laboratory under Contract DE-AC52-07NA27344. This work was funded by the Laboratory Directed Research and Development program of Lawrence Livermore National Laboratory (23-LW-043).

This work was carried out within the gA-EXPERiment and Theory (gA-EXPERT) collaboration—an international effort involving Kyungpook National University (KNU), the Institute for Basic Science (IBS), the Korea Research Institute of Standards and Science (KRISS), the University of Jyväskylä, and the Finnish Institute of Natural Resources (Luke)—dedicated to determining the effective axial-vector coupling constant, $g_{\rm A}^{\rm eff}$, through precision $\beta$-decay measurements and state-of-the-art nuclear theory comparisons.


\begin{thebibliography}{99}
\bibitem{Eng2017}
  J.~Engel and J.~Menendez, Rep. Prog. Phys. 60, 046301 (2017).

\bibitem{Eji2019}
  H.~Ejiri, J.~Suhonen, and K.~Zuber, Phys. Rep. 797, 1 (2019).

\bibitem{Ago2023} 
M.~Agostini, G.~Benato, J.~A. Detwiler, J.~Men{\' e}ndez, and F.~ Vissani, Rev. Mod. Phys. 95, 025002 (2023).

\bibitem{Suh2007} 
J.~Suhonen, \textit{From Nucleons to Nucleus: Concepts of Microscopic Nuclear Theory} (Springer, Berlin, 2007).

\bibitem{Beh1982} 
H.~Behrens and W.~B\"uhring, \textit{Electron Radial Wave Functions and Nuclear Beta-decay (International Series of Monographs on Physics)} (Clarendon Press, Oxford, 1982).

\bibitem{Haa2016} 
  M.~Haaranen, P.~C. Srivastava, and J.~Suhonen, Phys. Rev. C 93, 034308 (2016).

\bibitem{Haa2017} 
  M.~Haaranen, J.~Kotila, and J.~Suhonen, Phys. Rev. C 95, 024327 (2017).

\bibitem{Suh2017}
  J.~Suhonen, Front. Phys. 5, 55 (2017).

\bibitem{Suh2019} 
J.~Suhonen and J.~Kostensalo, Front. Phys. 7, 29 (2019).

\bibitem{Bod2020}
  L.~Bodenstein-Dresler \textit{et al.} (The COBRA Collaboration), Phys. Lett. B 800, 135092 (2020).

\bibitem{Kos2021}
 J.~Kostensalo, J.~Suhonen, J.~Volkmer, S.~Zatschler, and K.~Zuber, Phys. Lett. B 822, 136652 (2021).

\bibitem{Kos2023} 
J.~Kostensalo, E.~Lisi, A.~Marrone, and J.~Suhonen, Phys. Rev. C 107, 055502 (2023).

\bibitem{Ban2024} 
I. Bandac \textit{et al.}, Eur. Phys. J. C (2024) 84:1158.

\bibitem{Led2022} 
A.~F. Leder \textit{et al.}, Phys. Rev. Lett. 129, 232502 (2022).

\bibitem{Pag2024} 
L. Pagnanini \textit{et al.} (The ACCESS Collaboration), Phys. Rev. Lett. 133, 122501 (2024).

\bibitem{Kos2024} 
J.~Kostensalo, E.~Lisi, A.~Marrone, and J.~Suhonen, Phys. Rev. C 110, 045503 (2024).

\bibitem{Pau2023} 
M. Paulsen \textit{et al.}, Phys. Rev. C 110, 055503(2024). 

\bibitem{Kum2020}
  A.~Kumar, P.~C. Srivastava, J.~Kostensalo, and J.~Suhonen, Phys. Rev. C 101, 064304 (2020).

\bibitem{Kum2021}
  A.~Kumar, P.~C. Srivastava, and J.~Suhonen, Eur. Phys. J. A 57, 225 (2021).

\bibitem{Nitescu2023} O.~Nitescu, S.~Stoica, and F.~{\v S}imkovic, Phys. Rev. C 107 (2023) 025501.

\bibitem{Ram2024} 
M.~Ramalho and J.~Suhonen, Phys. Rev. C 109, 034321 (2024).

\bibitem{Ram2024c} 
M.~Ramalho, J.~Suhonen, A.~Neacsu, and S.~Stoica, Front. Phys. 12:1455778 (2024).

\bibitem{Ram2024b} 
M.~Ramalho and J.~Suhonen, Il Nuovo Cimento 47 C, 377 (2024).

\bibitem{Lis2004} A.~F. Lisetskiy, B.~A. Brown, M.~Horoi, and H.~Grawe, Phys. Rev. C 70, 0444314 (2004).

\bibitem{Mac1990} H.~Mach \textit{et al.}, Phys. Rev. C 41, 226 (1990).

\bibitem{MMC2024} J.~W.~Song, S.~G.~Kim, H.~S.~Kim, H.~J.~Kim, and M.~K.~Lee, J. Low Temp. Phys. 216, 436 (2024).

\bibitem{AgEr2025} J.~W.~Song, Y.~C.~Cho, H.~J.~Kim, and M.~K.~Lee, J. Low Temp. Phys. 218, 110 (2025).


\bibitem{kneading2016} M. P. Croce, A. S. Hoover, M. W. Rabin, E. M. Bond, L. E. Wolfsberg, D. R. Schmidt, and J. N. Ullom, J. Low Temp. Phys. 184, 938 (2016).

\bibitem{kneading2015} A. S. Hoover et al., Anal. Chem. 87, 3996 (2015).

\bibitem{SRS560} Stanford Research Systems, SR560 Low-Noise Preamplifier, Rev. 2.8 (Stanford Research Systems, Sunnyvale, CA, 2006), https://www.thinksrs.com/products/sr560.html


\bibitem{trap1994} V. ~T. ~Jordanov, G. ~F. ~Knoll, A. ~C. ~Huber, and J. ~A. ~Pantazis, Nucl. ~Instrum. Methods Phys. Res., Sect. A 353, 261 (1994).

\bibitem{pile2020} A. ~R. ~L. ~Kavner, D. ~Lee, S. ~T. ~P. ~Boyd, S. ~Friedrich, I. ~Jovanovic, and G. ~B. Kim, J. Low Temp. Phys. 209, 1070 (2022).


\bibitem{mat2016b} The MathWorks, Inc., MATLAB, version R2016b (The MathWorks, Inc., Natick, MA, 2016), https://www.mathworks.com/products/matlab.html

\bibitem{xray2003} R. ~D. ~Deslattes, E. ~G. ~Kessler Jr., P. ~Indelicato, L. ~De Billy, E. ~Lindroth, and J. ~Anton, Rev. Mod. Phys. 75, 35 (2003).

\bibitem{khazov2011nds} Y. ~Khazov, A. ~Rodionov, and F. ~G. ~Kondev, Nucl. Data Sheets \textbf{112}, 855 (2011).

\bibitem{LLBL2009x}A. C. Thompson et al., X-Ray Data Booklet (Center for X-Ray Optics and Advanced Light Source, Lawrence Berkeley National Laboratory, Berkeley, CA, 2009).

\end{thebibliography}
\end{document}